# Rich-clubness test: how to determine whether a complex network has or doesn't have a rich-club?

Alessandro Muscoloni[1,2] and Carlo Vittorio Cannistraci[1,2,3]

[1]Biomedical Cybernetics Group, Biotechnology Center (BIOTEC), [2]Center for Molecular and Cellular Bioengineering (CMCB), [3]Department of Physics, Technische Universität Dresden, Tatzberg 47/49, 01307 Dresden, Germany

Corresponding authors: Carlo Vittorio Cannistraci (kalokagathos.agon@gmail.com) and Alessandro Muscoloni (alessandro.muscoloni@gmail.com).

## Abstract

The rich-club concept has been introduced in order to characterize the presence of a cohort of nodes with a large number of links (rich nodes) that tend to be well connected between each other, creating a tight group (club). Rich-clubness defines the extent to which a network displays a topological organization characterized by the presence of a node rich-club. It is crucial for the investigation of internal organization and function of networks arising in systems of disparate fields such as transportation, social, communication and neuroscience. Different methods have been proposed for assessing the rich-clubness and various null-models have been adopted for performing statistical tests. However, a procedure that assigns a unique value of rich-clubness significance to a given network is still missing. Our solution to this problem grows on the basis of three new pillars. We introduce: i) a null-model characterized by a lower rich-club coefficient; ii) a fair strategy to normalize the level of rich-clubness of a network in respect to the null-model; iii) a statistical test that, exploiting the maximum deviation of the normalized rich-club coefficient attributes a unique p-value of rich-clubness to a given network. In conclusion, this study proposes the first attempt to quantify, using a unique measure, whether a network presents a *significant* rich-club topological organization. The general impact of our study on engineering and science is that simulations investigating how the functional performance of a network is changing in relation to rich-clubness might be more easily tuned controlling one unique value: the proposed rich-clubness measure.

# 1. Introduction

The rich-club concept has been introduced in 2004 by Zhou et al. [1], while studying the Internet topology at the autonomous system level, in order to characterize the presence of a cohort of nodes with a large number of links (rich nodes) that tend to be well connected between each other, creating a tight group (club). Consequently, the main motivation of the present study arises from the need to address a practical issue encountered in our research. While we were conducting a study on the latent geometry of complex networks [2], we had the need to perform a quantitative evaluation of the rich-clubness of synthetic networks created by a generative model, the Popularity-Similarity-Optimization (PSO) model [3]. The problem is that the PSO generative model consists of many parameters, whose combinations can generate networks of diverse topology and the question was how to test and quantify their level of rich-clubness. Unlikely, we did not find in literature a statistical test that was univocally assigning a significance value (p-value) for rich-clubness to a given network. However, the literature is rich of interesting studies and we will summarize below the state of the art.

The first method introduced for studying the rich-club property is the rich-club coefficient [1], defined for each degree $k$ as the density of the subnetwork composed of the nodes whose degree is larger than $k$:

$$\varphi(k) = \frac{2E_{>k}}{N_{>k}(N_{>k} - 1)}$$

where $E_{>k}$ is the number of edges between the $N_{>k}$ nodes whose degree is larger than $k$.

In 2006 Colizza et al. [4] argued that, being the rich-club coefficient a monotonically increasing function even for uncorrelated networks, a simple inspection of its trend is potentially misleading in the characterization of the rich-clubness. Therefore they suggested a normalization of the rich-club coefficient according to the formula:

$$\rho(k) = \frac{\varphi(k)}{\varphi_{rand}(k)}$$

where $\varphi_{rand}(k)$ is the rich-club coefficient computed in an appropriate null-model with the same degree distribution as the original network. A normalized rich-club coefficient greater than 1 shows the presence of the rich-club phenomenon with respect to the null case. The null-model adopted by Colizza et al. [4] and in further studies [5]–[9] is generated according to an iterative rewiring procedure proposed by Maslov and Sneppen (MS) for protein networks [10]. At each step two links are randomly selected and one endpoint of the first is exchanged with one endpoint of the second, provided that the links that would be created do not already exist. The degree of the nodes is automatically preserved and the appearance of multiple edges is avoided. The use of the MS procedure has been suggested also by Milo et al. [11], since it is fast and tends to a uniform sampling of random networks when enough iterations are performed. It has been compared with the configuration model [12], which is fast but suffers of non-uniform sampling. Furthermore, MS has been compared also with the go-with-the-winners algorithm [13], which is designed for a uniform sampling but its slowness prevents the use for large-scale networks.

In 2009 Zlatic et al. [14] addressed the problem of randomization of dense networks, where using the MS procedure the space of possible randomizations collapses and the null-model is close to the original network, making the assessment of rich-clubness hard. They introduced a generalization of the MS algorithm in which the network is considered as weighted, multiple edges are therefore allowed during the rewiring and the strength of the nodes is preserved rather than the degree. Accordingly, the rich-club coefficient is also generalized using the node strength in place of the node degree. This solution is able to generate a null-model with a rich-club coefficient close to the MS procedure for sparse networks and lower than the MS procedure for increasing network density. Other studies have extended the rich-club coefficient also to weighted networks [15], [16].

A methodological change in the characterization of the rich-clubness has been presented in 2008 by Jiang et al. [6], arguing the need of a statistical test for the assessment. A population of random networks is generated according to the null-model, the rich-club coefficient is computed for all of them and for the original network, lastly a one-sided p-value is assigned to each $k$ as the percentage of $\varphi_{rand}(k)$ that are greater or equal than $\varphi(k)$. The same procedure has been applied in further studies while studying the rich-clubness in brain networks [7]–[9], using even the Bonferroni correction [8] or the false discovery rate correction [9] for multiple comparisons over the range of degree $k$. After this summary of the state of art for the characterization of the rich-clubness, the main points of weakness will be discussed and the new approach will be introduced. However, it is clear that the current state of the art does not provide any unique measure of rich-clubness that quantifies whether a network present a *significant* rich-club topological organization.

## 2. Methods

### 2.1. Null-model

The first limitation of the state of the art approach for the assessment of the rich-clubness lies at the beginning of the procedure: the null-model. Although the MS procedure [10] is fast and tends to a uniform sampling of random networks [11], it does not always represent a suitable choice for the evaluation of the rich-clubness. Fig. 1 shows the rich-club coefficient for synthetic networks generated by the PSO model using several parameter combinations, compared with the rich-club coefficient computed in the MS null-model (mean over 1000 random networks). The figure highlights that for most of the parameter combinations, in particular for large network size (*N*) and average node degree (2*m*), the synthetic networks present fully connected ($\varphi(k) = 1$) *k*-subnetworks for high degrees, which might suggest the presence of rich-clubness. However, it can be noticed that also the MS null-model is characterized by a rich-club coefficient very close to the one of the original network. This represents a relevant limitation of the MS procedure when adopted in this context. Networks could be classified as non-rich-club not because the rich nodes do not form a club, but because the rich nodes form a club also in the null-model. The first target was therefore to design a new null-model characterized by a lower rich-club coefficient.

Zhou et al. [17] showed that, for a given degree distribution, the rich-club coefficient is very sensitive to changes in the degree-degree correlation (assortativity) of the network. Furthermore, we noticed that the MS procedure could be generalized so that the two edges to swap can be sampled with any given probability (see Algorithm 1.A), which is uniform in the particular case of the MS null-model (see Algorithm 1.B). Following these two hints, we conceived a new null-model, which for simplicity (in order to easily differentiate from the Maslov-Sneppen) we named Cannistraci-Muscoloni (CM). The CM model is a generalization of the MS model where the two probabilities of the two sampled swapping edges are defined separately in function of the adjacent nodes' degree and have an 'inverted tendency'. For instance, one edge is sampled with probability directly proportional to the product of the degrees of the adjacent nodes, whereas the other edge with probability inversely proportional (see Algorithm 1.C). This rule tends to sample one edge connecting high degree nodes and one edge connecting low degree nodes. When two endpoints are swapped, the procedure breaks links between high-high and low-low degree nodes and establishes links between high-low degree nodes, decreasing the assortativity of the network and in turn also the rich-club coefficient. Fig. 1, together with the rich-club coefficient of the original network and of the MS null-model, compares also the one of the CM null-model (mean over 1000 random networks), confirming the theoretical expectations of the previous discussion. For sake of completeness, Suppl. Fig. 1 shows the comparison with a further variant of the CM null-model in which the sampling probabilities are proportional to the sum of the degrees of the connected nodes. However, the adoption of the sum resulted less effective than the product.

## 2.2. Normalization

The second limitation of the state of the art approach for the assessment of the rich-clubness is that it does not assign a unique p-value to the network under investigation. In order to do it using a robust statistical test, we considered important to focus on the maximum deviation of the rich-club coefficient from the null-model. However, we noticed that the previous normalization using the ratio is not suitable for this purpose, since it generates a hyperbolic effect that emphasizes a higher deviation at lower degrees and this could lead to the consideration of the wrong maximum. Fig. 2 presents both a theoretical plot and a real example in a PSO network, visually highlighting the point discussed. In order to solve this problem, we devised a different normalization of the rich-club coefficient according to the formula:

$$\rho(k) = \varphi(k) - \varphi_{rand}(k)$$

where $\varphi_{rand}(k)$ is the rich-club coefficient computed in the null-model, in this case the CM. The normalized coefficient assumes values in the range [-1, 1] and a value of 0 represents the reference case in which the rich-club coefficient is equal to the one in the null-model. The adoption of the difference rather than the ratio correctly preserves the distances between the rich-club coefficients and ensures the detection of the actual maximum deviation from randomness (Fig. 2).

## 2.3. Statistical test

Once introduced a suitable null-model and a proper normalization for the rich-club coefficient, we will now describe the statistical test that we proposed in this study in order to assign a unique p-value for rich-clubness. For a given network in input, the procedure is as follows:

(1) A population of random networks are generated using the CM null-model proposed in this study (in our simulations we used 1000 repetitions).
(2) The rich-club coefficient of the network under investigation is computed for each degree and then normalized (using the difference) by the mean coefficient of all the random networks.
(3) The rich-club coefficient of every random network is also computed for each degree and normalized by the mean coefficient of all the random networks.
(4) The maximum value of the normalized rich-club coefficient (peak of deviation from the mean null-model) is computed both for the considered network (observed peak) and for the population of random networks (null distribution of peaks). In practice, the ensemble of all the random peaks (one for each random network generated by the CM model) creates a null distribution of random peaks.

(5) A one-sided p-value is computed as the percentage of random peaks greater or equal than the observed peak.

An explanatory plot of the statistical test for rich-clubness is shown in Fig. 4, the results will be discussed in the next section.

## 3. Results and Discussion

The first example of application of the statistical test for rich-clubness proposed in this study is presented in Fig. 3-5. Fig. 3A shows two PSO networks generated with different parameters, in particular they have the same number of nodes and edges, but the PSO network 2 has lower temperature, leading to a more hierarchical and clustered structure, as can be seen in its visual representation. Fig 3B reports for the two networks the rich-club coefficient (non-normalized) of the original network and the one of the CM null-model (mean over 1000 random networks). The plot suggests that both the PSO networks might have a dense subnetwork of rich nodes, however, intuitively, the PSO network 1 presents a region of high degrees with a greater deviation from the null-model. This is confirmed in Fig. 4, which highlights an explanatory plot of the statistical test for rich-clubness for the two synthetic networks. The PSO network 1 shows a higher peak of normalized rich-club coefficient and only a percentage $p = 0.006$ of peaks in the population of random networks are greater or equal, underlining the presence of a significant rich-club in the network. The subnetwork corresponding to the observed peak is shown in Fig. 5 and consists of the five highest degree nodes, which are fully connected between each other. On the contrary, the PSO network 2 is characterized by a higher $p = 0.093$, which falls above the significance level of 0.05, although not far.

In order to assess the rich-clubness trend for several parameter combinations of the PSO model, we generated 10 synthetic networks for each combination of size $N = [100, 500, 1000]$, half of average degree $m = [2, 4, 6]$ and temperature $T = [0, 0.3, 0.6, 0.9]$, fixing the power-law degree distribution exponent $\gamma = 2.5$. For each PSO network we generated 1000 random networks using the null-model CM and the statistical test for rich-clubness has been performed. The p-values have been adjusted for multiple hypothesis testing (over the 10 networks) by the Bonferroni correction. Fig. 6 reports the average of the adjusted p-values for each different parameter combination, highlighting the range below the significance level of 0.05. The statistical test highlights that for most of the parameter combinations, in particular for $m = [4$,

6] and $N = [500, 1000]$ the networks present a significant rich-club, whereas for more sparse ($m = 2$) and small networks ($N = 100$) in general there is not a significant rich-club. This is in agreement with the network growing procedure explained by the PSO model. In fact, the high degree nodes are the first ones to be born in the network and they are expected to connect to around $m$ of the older nodes [3]. Therefore for higher $m$ the rich nodes have higher probability to create a club.

Finally, we apply the statistical test for rich-clubness to real networks. The networks considered are the same analysed by Colizza et al. [4] and represent different physical systems (see the Dataset section for further details): an Internet network (*AS200105*), a transportation network (*USAir500*), a scientific collaboration network (*condmat1998*) and a protein-protein interaction network (*DIPyeast*). Fig. 7 reports for each network the rich-club coefficient (non-normalized) of the original network and the one of the MS and CM null-models (mean over 1000 random networks). The advantage of the CM null-model is particularly evident in the network *AS200105*, where both the original network and the MS null-model have a very high rich-club coefficient for the upper half of the degree range, whereas the CM null-model reduces it in a consistent way. The results of the statistical test for rich-clubness are reported in Fig. 8, together with a comparison of the state of the art approach (MS null-model, normalization using the ratio and separate statistical test for each degree).

*AS200105* presents a significant rich-club according to our evaluation ($p < 0.001$), whereas the state of the art approach suggests that at none of the degrees the rich-club coefficient is significantly higher than the null-model. The rich-clubness of the Internet networks has been claimed in the first study led about this network property [1] and has been quite debated in further studies [4], [6], [17], without reaching a final agreement. Here we want to focus on the particular topology analysed and we believe that the lack of rich-clubness, supported by the state of the art approach and stated also by Colizza et al. [4], is a misleading result due to the unsuitability of the MS null-model. Indeed, as already shown in Fig. 7, both the original network and the MS null-model present fully connected ($\varphi(k) = 1$) $k$-subnetworks in the upper half of the degree range, compromising the detection of this very tight club of rich nodes after the normalization of the coefficient. We notice that this problem has been raised also by Zhou et al. [17].

*USAir500* presents a significant rich-club according to our evaluation ($p < 0.001$), a result that is supported also by the state of the art approach, suggesting that the rich-club coefficient is significantly higher than the null-model for most of the degree values ($p < 0.05$ with and without corrections), including the degree corresponding to the observed peak of deviation. In previous studies, a mild rich-club organization has been claimed for the U.S. airport network also by Colizza et al. [4] and Serrano [16]. The presence of tight interconnections between hubs has been observed even in the worldwide airport [18] and in India's airport network [19].

*condmat1998* presents a significant rich-club according to our evaluation ($p = 0.049$). The result is partially supported by the state of the art approach, suggesting that the rich-club coefficient is significantly higher than the null-model for most of the degree values ($p < 0.05$ with and without corrections), but not for the degree corresponding to the observed peak of deviation. The fact that scientists with more collaborations tend to form collaborative groups within specific research domains has been supported also in previous studies [4], [15], [16].

*DIPyeast* does not present a significant rich-club according to our evaluation ($p = 0.357$). Colizza et al. [4] reported the same result, commenting that proteins with a large number of interactions are presiding over different functions and thus, in general, are coordinating specific functional modules. The lack of rich-clubness in the protein-protein interaction networks of *S. cerevisiae* is confirmed also in further studies [6], [20]. The state of the art approach suggests that the rich-club coefficient is significantly higher than the null-model for a few low degree values ($p < 0.05$ with and without corrections). This result looks illogical and highlights a point of weakness of the procedure. It detected significant p-values only for the lowest degree values, whereas the rich-clubness is a network property that should characterize the rich nodes.

To conclude, we briefly summarize the main innovations introduced in this study. We designed the CM null-model which is able to generate random networks with a lower rich-club coefficient, more suitable for the characterization of the rich-clubness. In particular, it was able to solve the main point of weakness of the MS procedure, which can produce random networks with a rich-club coefficient as high as in the original network, bringing to misleading conclusions. We proposed a new statistical test which is the first to assign a unique p-value to a given network. The test focus on the maximum deviation of the rich-club coefficient of the observed network from the null-model, therefore we introduced a new normalization procedure using the difference rather than the ratio. The rationale is that the difference is an operator that fairly adjusts the RC-coefficient of the observed network, in fact the normalization function exerts the same correction regardless of the level of degree *k*. This new normalization favours the detection of the correct peak of deviation from the null-model. The test presents robustness since it exploits a population of random peaks of deviation from the average null-model that might occur also at different degree values. Contrarily to the previous statistical tests (which provided a p-value for each degree), our test does not produce any significant p-value for low-degree cohorts (which are, in our opinion, erroneously indicated as rich-clubs in previous literature). Indeed, detecting low-degree rich-clubs represents a paradox, since the rich-club property should characterize a subnetwork of rich nodes in comparison to the rest of the network. At last, we leave a point of investigation for future studies, underlining that our statistical test can be easily adapted for the characterization of the weighted rich-club coefficient as defined in [15], [16] and that the CM null-model is compatible with the procedure suggested by Zlatic et al. [14], which could help even further for the reduction of the rich-club coefficient while generating a null-model for dense networks.

We believe that the general impact of our study on engineering and science is in particular in large scale simulations and network design. In fact, simulations that investigate how the functional performance of a network is changing in relation to rich-clubness might be more easily tuned controlling one unique value that is the proposed rich-clubness measure.

## 4. Dataset

### 4.1. Synthetic networks generated using the PSO model

The synthetic networks used in this study have been created according to the PSO model [3], which describes how random geometric graphs grow in the hyperbolic space.

The model has four input parameters: $m > 0$, which defines the average node degree $\bar{k} = 2m$, $\beta \in (0, 1]$, defining the exponent $\gamma = 1 + 1/\beta$ of the power law degree distribution, $T \geq 0$, which controls the network clustering, and $\zeta = \sqrt{-K} > 0$, where $K$ is the curvature of the hyperbolic plane. The network clustering is maximized at $T = 0$, it decreases almost linearly for $T = [0,1)$ and it becomes asymptotically zero if $T > 1$.

Building a network of $N$ nodes on a hyperbolic plane with curvature $K = -1$ requires the following steps: (1) Initially the network is empty; (2) At time $i = 1, 2, ..., N$ a new node $i$ appears with radial coordinate $r_i = 2ln(i)$ and angular coordinate $\theta_i$ uniformly sampled from $[0,2\pi]$; all existing nodes $j < i$ increase their radial coordinates according to $r_j(i) = \beta r_j + (1 - \beta)r_i$ in order to simulate popularity fading; (3) The new node picks a randomly chosen existing node and connects to it with probability $p(h_{ij}) = 1/(1 + exp((h_{ij} - R_i)/2T))$, where $R_i = r_i - 2\ln\left[\frac{2T(1-e^{-(1-\beta)\ln(i)})}{\sin(T\pi)m(1-\beta)}\right]$ is the current radius of the hyperbolic disk, $h_{ij} = arccosh(\cosh r_i \cosh r_j - \sinh r_i \sinh r_j \cos \theta_{ij})$ is the hyperbolic distance between node $i$ and node $j$ and $\theta_{ij} = \pi - |\pi - |\theta_i - \theta_j||$ is the angle between these nodes. Step (3) is repeated until the new node $i$ is connected to $m$ nodes. (4) The growing process stops when N nodes have been introduced.

In this model networks evolve optimizing a trade-off between node popularity, abstracted by the radial coordinate, and similarity, represented by the angular coordinate, and they exhibit many common structural and dynamical characteristics of real networks.

### 4.2. Real networks

The four real networks considered in this study are the same analysed by Colizza et al. [4] and have been downloaded at the website https://sites.google.com/site/cxnets/data. The networks have been transformed into undirected, unweighted, without self-loops and only the largest connected component has been considered. A basic description, the data source and number of nodes $N$ and edges $E$ are now reported.

*AS200105* ($N = 11174$, $E = 23409$): network of the Internet at the Autonomous System level from data collected by the Oregon Route Views Project (http://www.routeviews.org/) in May 2001, where nodes represent Internet service providers and edges connections among them.

*USAir500* ($N = 500$, $E = 2980$): network as obtained by considering the 500 US airports with the largest amount of traffic from publicly available data. Nodes represent US airports and edges represent air travel connections among them [21].

*condmat1998* ($N = 12722$, $E = 39967$): network extracted from the electronic database e-print archive in the area of condensed matter physics (http://xxx.lanl.gov/archive/cond-mat/) from 1995 to 1998, in which nodes represent scientists and a connection exists if they co-authored at least one paper contained in the archive [22], [23].

*DIPyeast* ($N = 4626$, $E = 14801$): network of the yeast *Saccharomyces cerevisae* extracted with different experimental techniques and collected at the Database of Interacting Proteins (http://dip.doe-mbi.ucla.edu/) [24].

**Hardware and software**

MATLAB code has been used for all the simulations, carried out partly on a workstation under Windows 8.1 Pro with 512 GB of RAM and 2 Intel(R) Xenon(R) CPU E5-2687W v3 processors with 3.10 GHz, and partly in the ZIH-Cluster Taurus of the TU Dresden.

**Funding**

Work in the CVC laboratory was supported by the independent research group leader starting grant of the Technische Universität Dresden. AM was partially supported by the funding provided by the Free State of Saxony in accordance with the Saxon Scholarship Program

Regulation, awarded by the Studentenwerk Dresden based on the recommendation of the board of the Graduate Academy of TU Dresden.

**Acknowledgements**

We thank Alexander Mestiashvili and the BIOTEC System Administrators for their IT support, Claudia Matthes for the administrative assistance and the Centre for Information Services and High Performance Computing (ZIH) of the TUD. We thank the anonymous reviewer that asked to evaluate the level of Rich-clubness in PSO models, which occurred in another parallel study we have been conducting on latent geometry of complex networks.

**Author contributions**

CVC conceived the new generalized null-model and the new sampling probabilities. AM devised the new normalization based on the difference. CVC invented the new statistical test. AM implemented the code and performed the computational analysis. Both the authors analysed and interpreted the results. AM wrote the draft of the article according to CVC suggestions and CVC corrected and improved it to arrive to the final draft. CVC designed the figures and AM realized them. CVC planned, directed and supervised the study.

**Competing interests**

The authors declare no competing financial interests.

**Algorithm 1. Generative procedure and sampling probabilities for the null-models.**

**A. Generalized Null-model procedure (proposed by Cannistraci and Muscoloni)**

INPUT: original network, iterations *M*
OUTPUT: random network

*count* = 0;
while *count* < *M*
    sample one edge (*a,b*) with probabilities *p1*
    sample one edge (*c,d*) with probabilities *p2*

    if the nodes (*a,b,c,d*) are all different
        choose randomly between alternatives [I] and [II]
        if [I] and both edges (*a,d*),(*b,c*) do not already exist
            swap edges (*a,b*),(*c,d*) to (*a,d*),(*b,c*)
            *count* = *count* + 1
        if [II] and both edges (*a,c*),(*b,d*) do not already exist
            swap edges (*a,b*),(*c,d*) to (*a,c*),(*b,d*)
            *count* = *count* + 1

Note: in this study we fixed *M* = 10\**E*, where *E* is the number of edges in the network.

**B. Maslov-Sneppen (MS) sampling probabilities**

Associate to every edge $(i,j)$ a probability:
$$p1(i,j) = p2(i,j) = \frac{1}{E}$$
Therefore the two edges are sampled with uniform probability.

**C. Cannistraci-Muscoloni (CM) sampling probabilities**

Compute for each node $i$ the degree $k(i)$.
Create a vector $w1$ that associates to every edge $(i,j)$ a weight:
$w1(i,j) = k(i) * k(j)$
Create a vector $w2$ that associates to every edge $(i,j)$ the weights of $w1$ reversed:
$w2(i,j) = |w1(i,j) - \max(w1) - \min(w1)|$
Compute for each edge $(i,j)$ the probabilities:
$$p1(i,j) = \frac{w1(i,j)}{\text{sum}(w1)}$$
$$p2(i,j) = \frac{w2(i,j)}{\text{sum}(w2)}$$
Therefore one edge is sampled with probability directly proportional to the product of the nodes degree and one edge with probability inversely proportional to the product of the nodes degree. Note: when two edges are successfully swapped, the probabilities have to be updated.

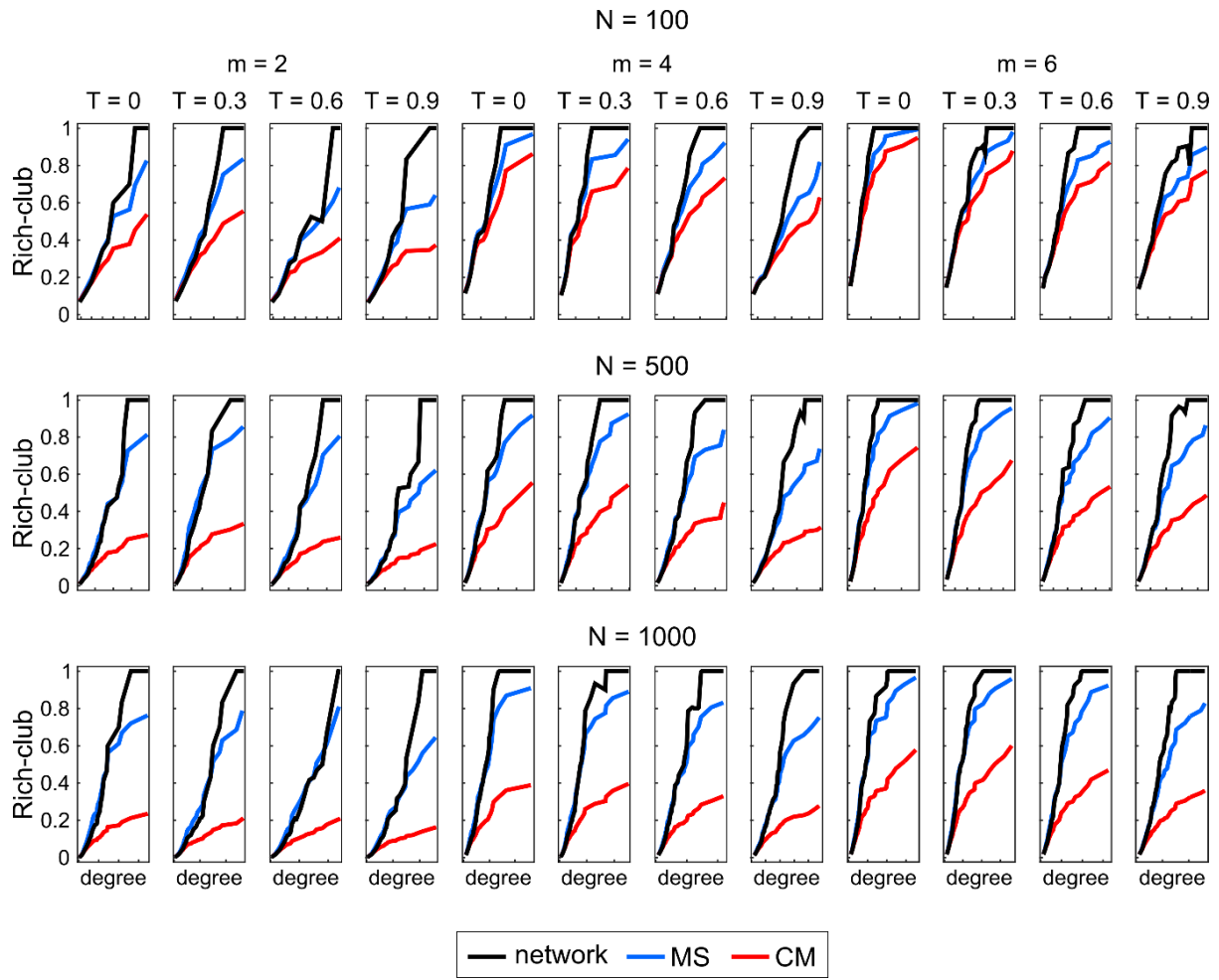

**Figure 1. Comparison of null-models for the rich-club coefficient.**

We created a synthetic network for each combination of tuneable parameters of the Popularity-Similarity-Optimization (PSO) model (size $N$, half of average degree $m$, temperature $T$), fixing the power-law degree distribution exponent $\gamma = 2.5$. For each PSO network we generated 1000 random networks using the two null-models discussed in this study: Maslov-Sneppen (MS) and Cannistraci-Muscoloni (CM). The plots report, for each different parameter combination, the rich-club coefficient (non-normalized) of the PSO network and the mean rich-club coefficient for each null-model, averaged over the 1000 repetitions.

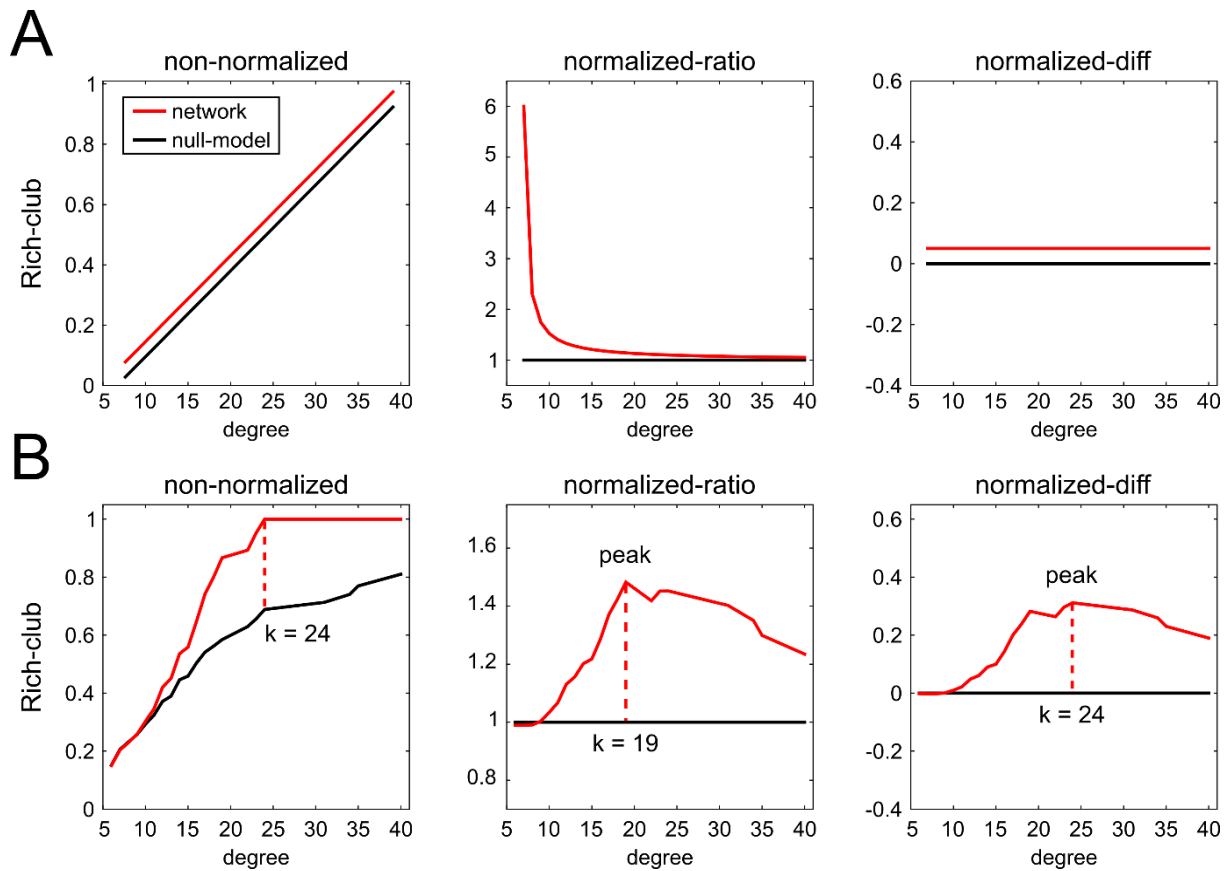

**Figure 2. Comparison of normalizations for the rich-club coefficient.**
(**A**) Theoretical plots of the rich-club coefficient (non-normalized and normalized using the ratio or the difference) for the original network and the null-model. In presence of a constant deviation for increasing degree (left panel) between the rich-club non-normalized coefficient of the original network and the one of the null-model, the normalization using the difference (right panel) preserves the distances, whereas the normalization using the ratio (central panel) introduces a hyperbolic effect, showing a higher deviation at lower degrees. (**B**) Non-normalized rich-club coefficients (left panel) for a PSO network ($N = 100$, $m = 6$, $T = 0.6$, $\gamma = 2.5$) and mean rich-club coefficient for the CM null-model (1000 repetitions). The normalization using the ratio (central panel) gives higher emphasis at lower degrees leading to the detection of the wrong peak (k=19) of deviation from randomness. For each panel the value $k$ of the degree corresponding to the peak is reported. It is evident that the normalization using the difference (right panel) ensures the detection of the correct peak (k=24) for which emerges the rich-club.

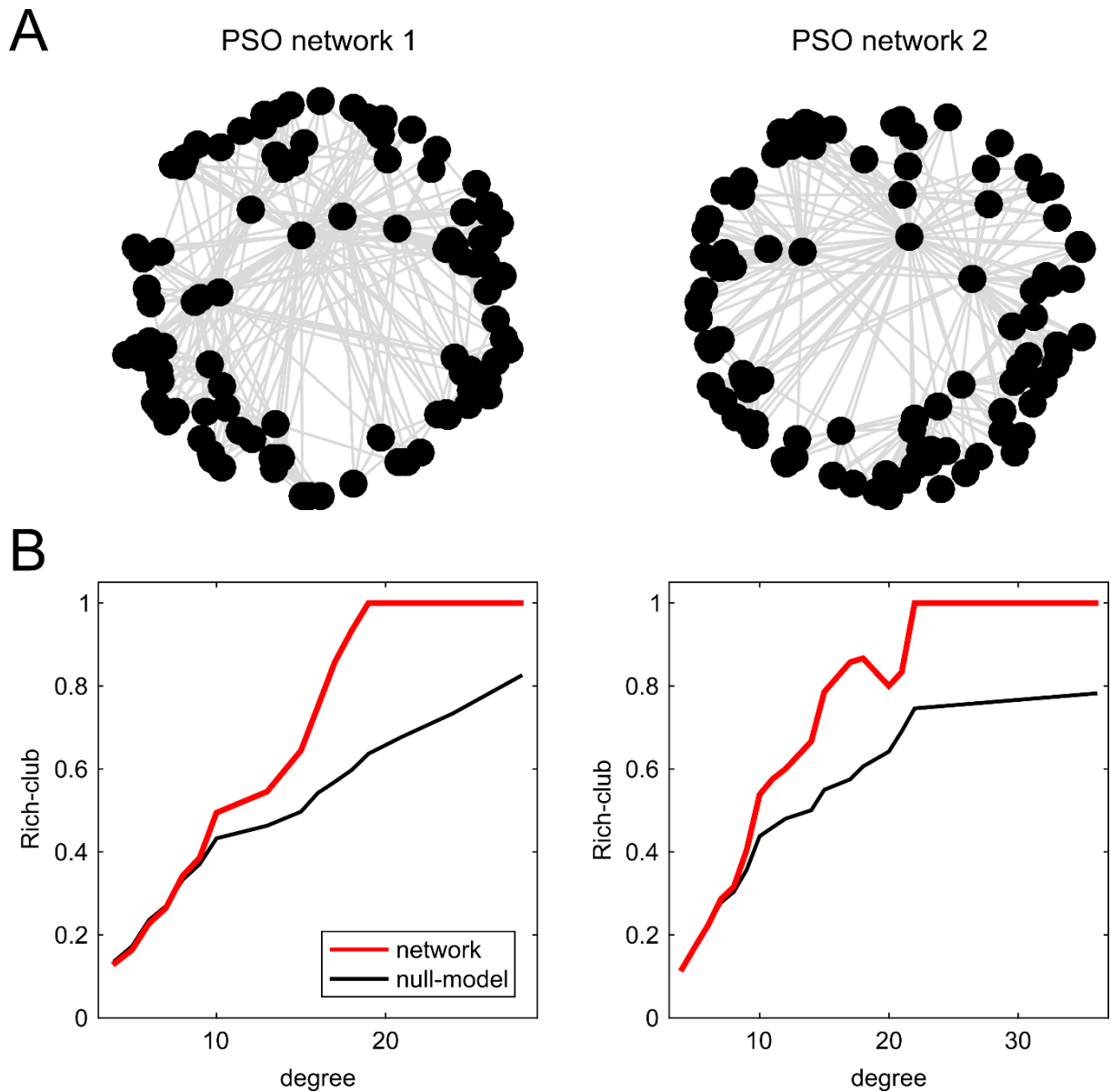

**Figure 3. Rich-club coefficient on two PSO networks.**

(**A**) Visual representation of two PSO networks generated with parameters: (1) $N = 100$, $m = 4$, $T = 0.3$, $\gamma = 2.5$; (2) $N = 100$, $m = 4$, $T = 0$, $\gamma = 2.5$. (**B**) The plots report for the two PSO networks the rich-club coefficient (non-normalized) of the original network and the mean rich-club coefficient of the CM null-model (1000 repetitions).

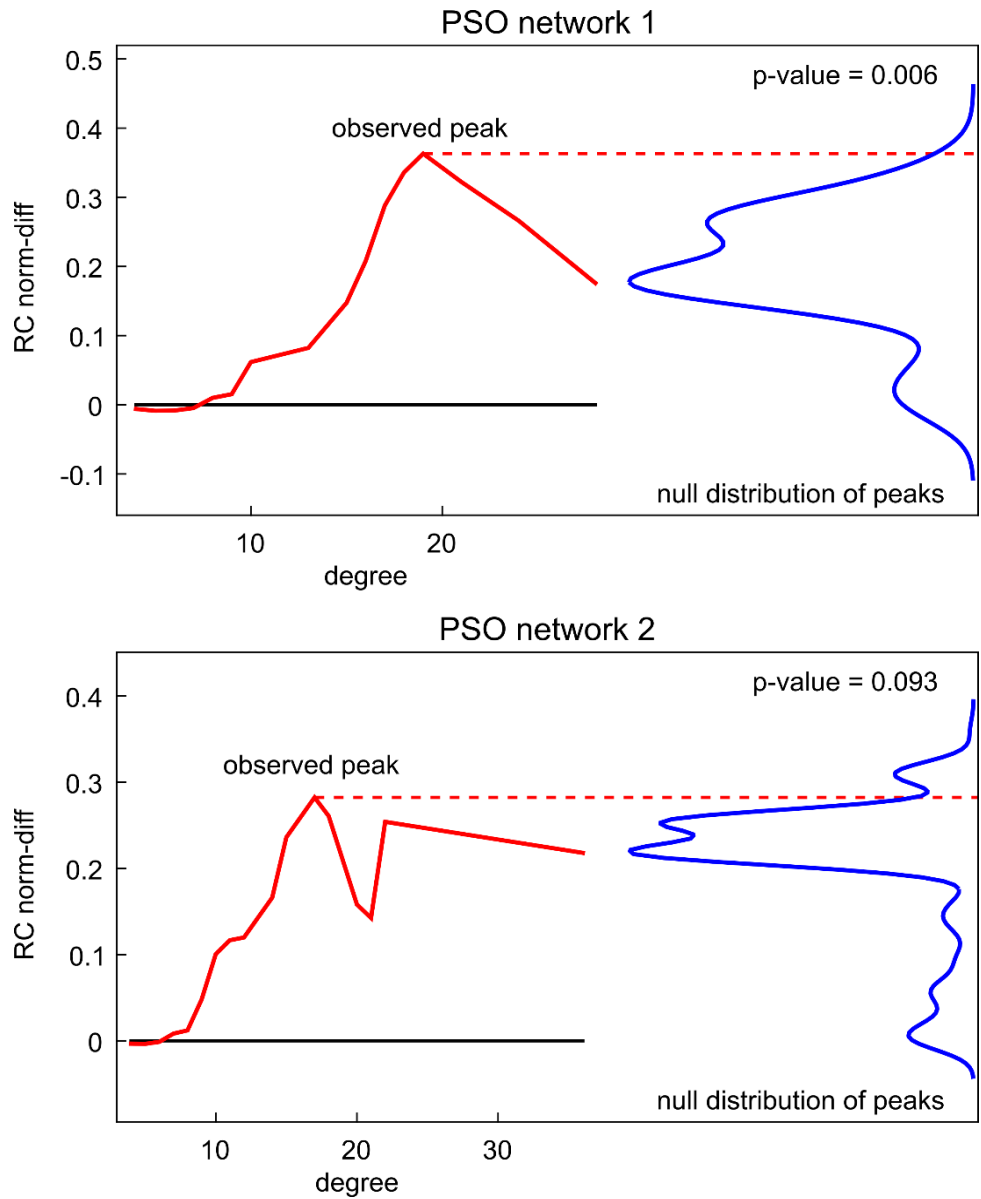

**Figure 4. Rich-clubness statistical test on two PSO networks.**

Explanatory plot of the statistical test for rich-clubness, reported for the two PSO networks shown in Fig. 3. (1) A population of 1000 random networks are generated using the CM null-model proposed in this study. (2) The rich-club coefficient of the considered network is computed for each degree and then normalized (using the difference) by the mean coefficient of the random networks. The normalized coefficient is shown in red, whereas the black ground line indicates the reference case in which the rich-club coefficient is equal to the mean coefficient of the null-model. (3) The rich-club coefficient of every random network is also computed for each degree and normalized. (4) The maximum value of the normalized rich-club coefficient (peak of deviation from the mean null-model) is computed both for the considered network (observed peak) and for the population of random networks (null distribution of peaks). The red line on the left reports the normalized RC coefficient curve obtained at different degrees for the real network, notice that to preserve clarity the 1000 normalized RC curves for the random networks are not reported. However, the blue line on the right shows the null distribution of random peaks generated by the random networks obtained by the CM null-model. The dashed red line represents the projection of the observed peak to the null distribution of random peaks. (5) A one-sided p-value is computed as the percentage of random peaks greater or equal than the observed peak.

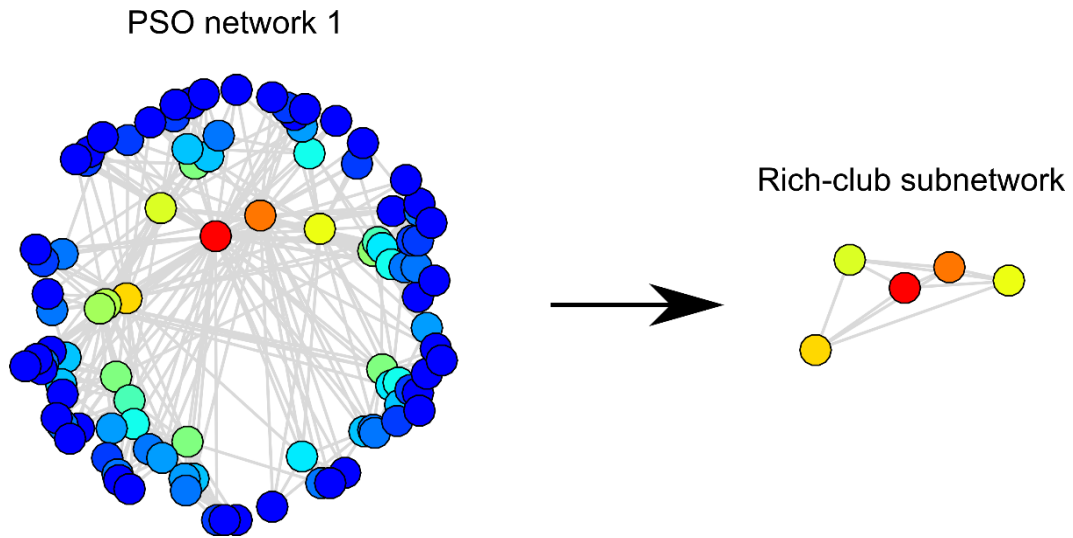

**Figure 5. PSO subnetwork corresponding to a significant rich-club.**
On the left, a visual representation of the same PSO network 1 presented in Fig. 3 and that showed a significant p-value for rich-clubness in Fig. 4. The nodes are coloured by increasing log-degree using a cold-to-warm colour map. On the right, the rich-club, i.e. the subnetwork composed by the nodes with degree larger than the degree of the observed peak (maximum deviation from the CM null-model) in the normalized rich-club coefficient.

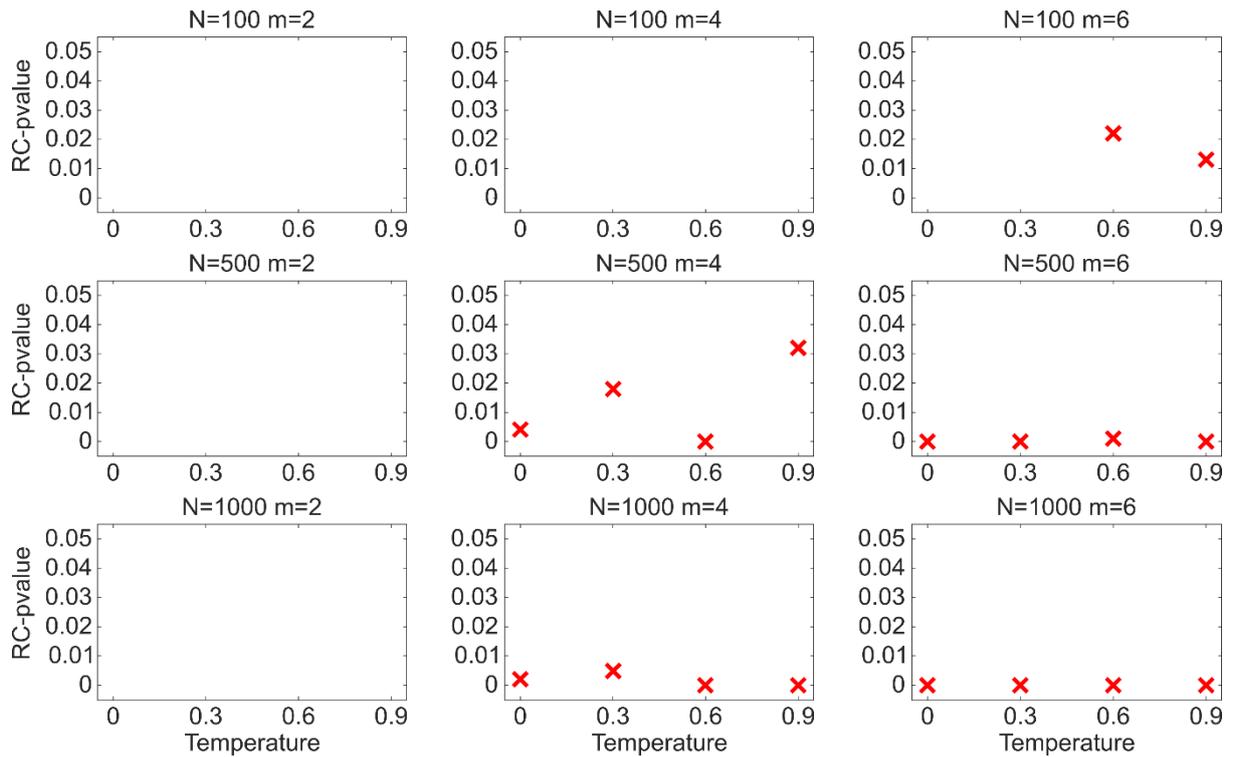

**Figure 6. P-values of the statistical test for rich-clubness on PSO networks.**

We created 10 synthetic networks for each combination of tuneable parameters of the PSO model (size $N$, half of average degree $m$, temperature $T$), fixing the power-law degree distribution exponent $\gamma = 2.5$. For each PSO network we generated 1000 random networks using the CM null-model and the statistical test for rich-clubness has been performed. The p-values have been adjusted for multiple hypothesis testing (over the 10 networks) by the Bonferroni correction. The plots report, for each different parameter combination, the average of the adjusted p-values, highlighting the range below the significance level of 0.05.

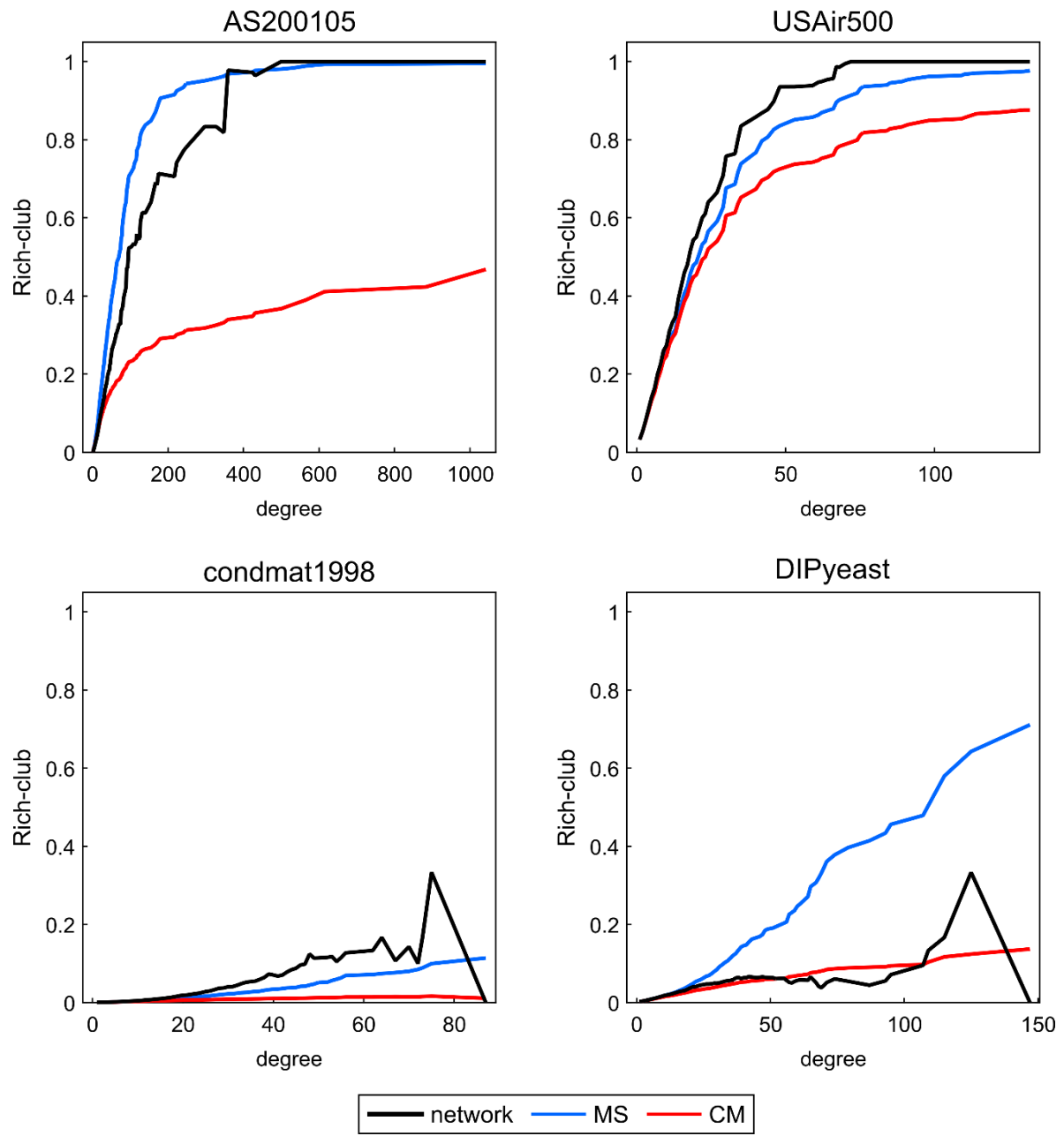

**Figure 7. Rich-club coefficient on real networks.**

The plots report, for four real networks, the rich-club coefficient (non-normalized) of the original network and the mean rich-club coefficient of the null-models using MS and CM (1000 repetitions). The networks represent different physical systems (see the Dataset section for further details): Internet network (*AS200105*), transportation network (*USAir500*), scientific collaboration network (*condmat1998*) and protein-protein interaction network (*DIPyeast*).

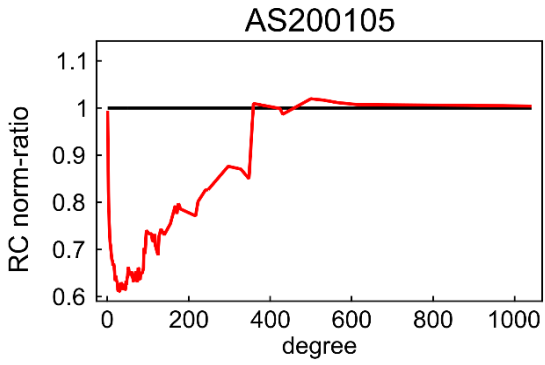
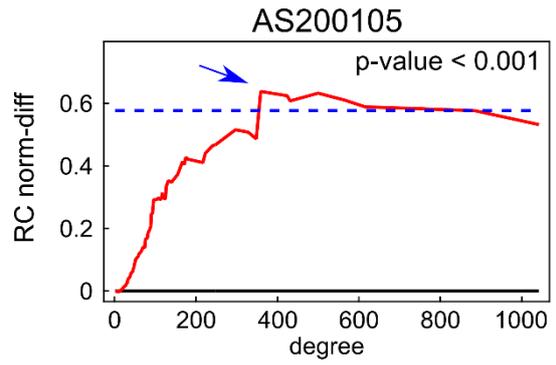
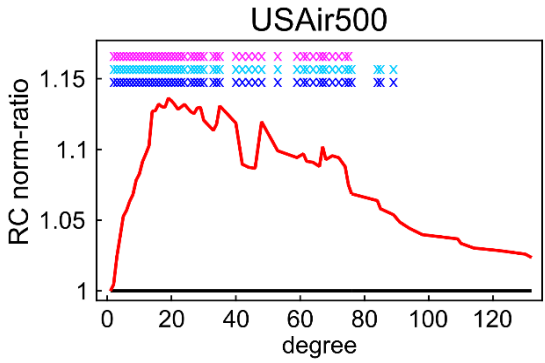
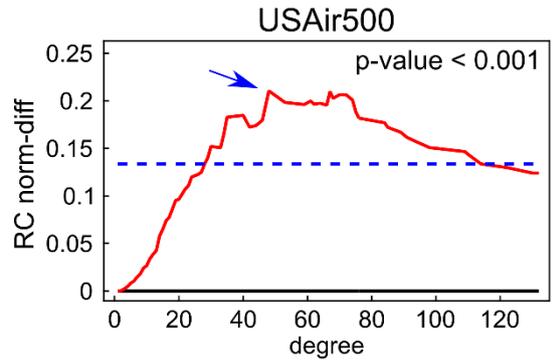
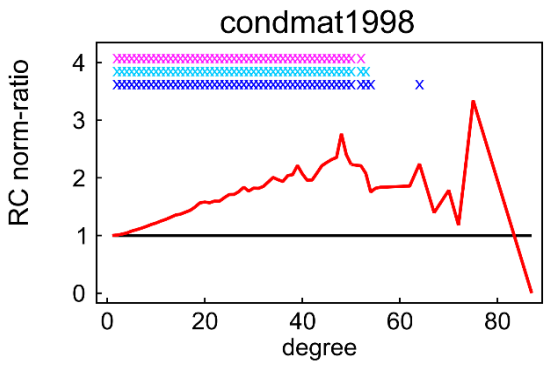
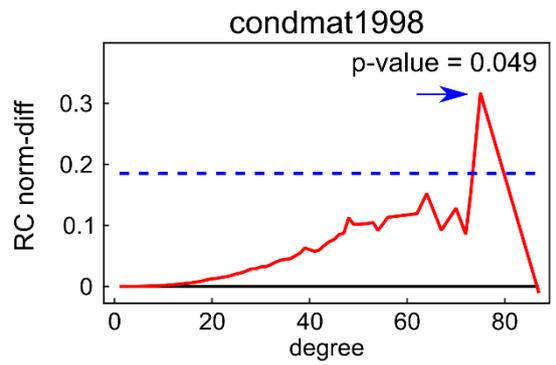
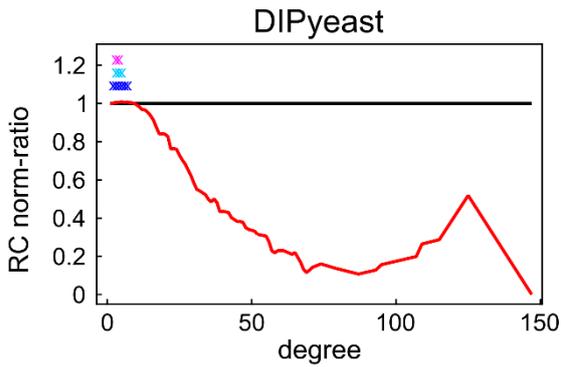
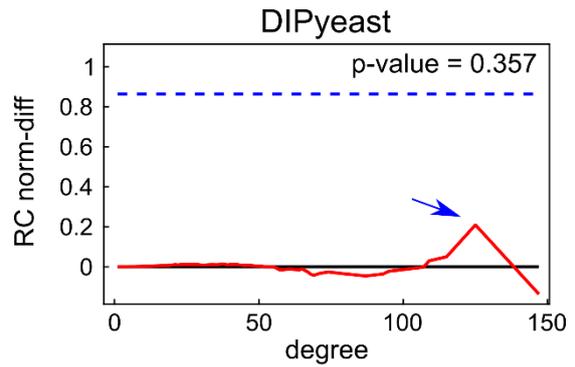

**Figure 8. State of the art approach and statistical test for rich-clubness on real networks.**
For the four real networks introduced in Fig 7, the figure shows both the state of the art approach for measuring the rich-clubness (left column) and the statistical test proposed in this study (right column). The plots on the left column present the rich-club coefficient normalized using the ratio by the mean coefficient of the MS null-model (1000 repetitions). For each degree, a p-value is computed as the percentage of rich-club coefficients of the random networks that are greater or equal than the rich-club coefficient observed at that degree in the original network. If the p-value is within the significance level of 0.05, an 'X' is reported. P-values without correction and with Benjamini-Hochberg or Bonferroni correction for multiple hypothesis testing over the degree range are considered (reported with different colours). The plots on the right column present the rich-club coefficient normalized using the difference by the mean coefficient of the CM null-model (1000 repetitions). The p-value of the statistical test for rich-clubness is reported and the corresponding peak of the normalized rich-club coefficient is indicated by a blue arrow. The dashed blue line represents the confidence level below which it falls more than 5% of the random peaks of the sampled null distribution of random peaks.

# SUPPLEMENTARY

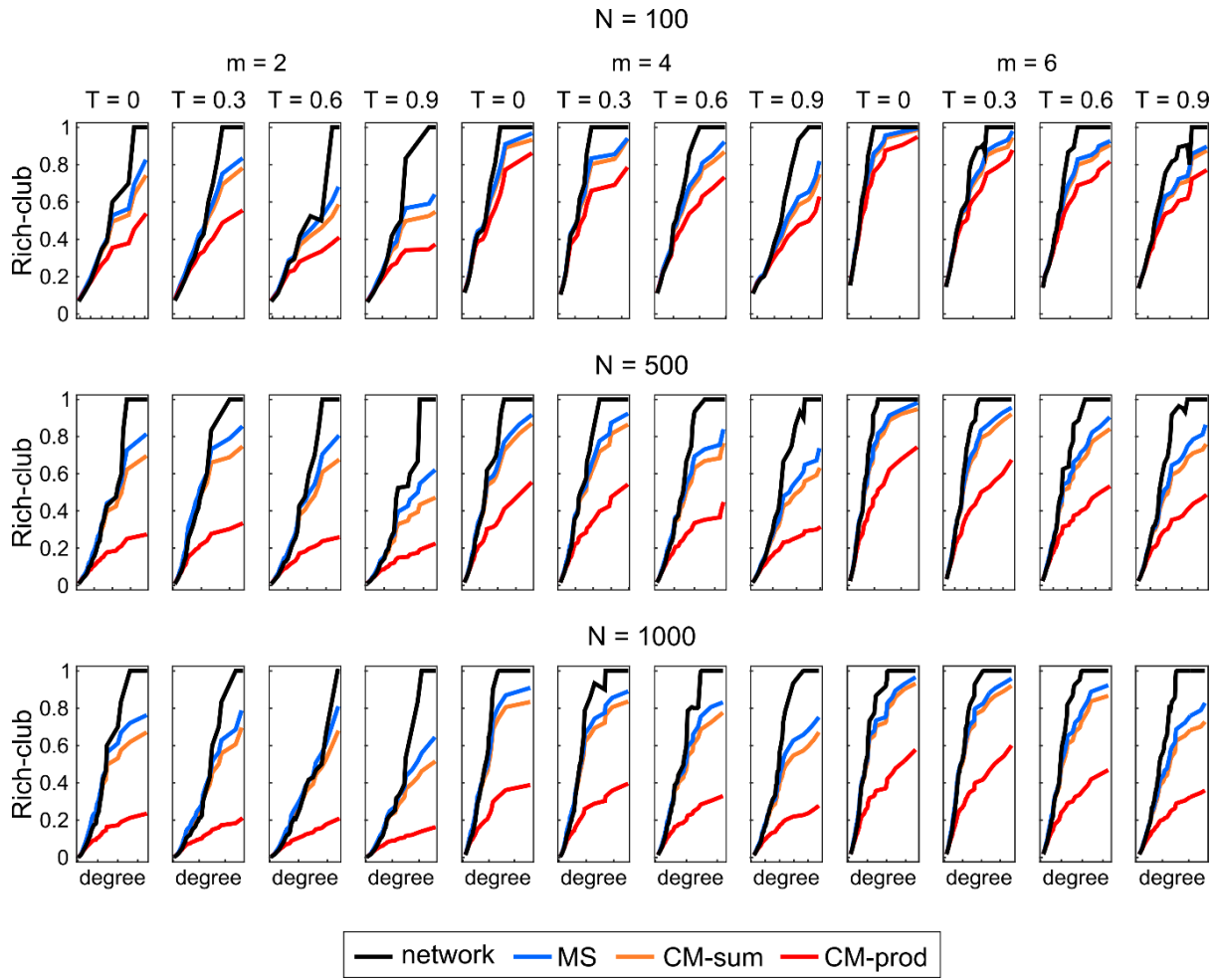

**Figure 1. Comparison of null-models for the rich-club coefficient.**
We created a synthetic network for each combination of tunable parameters of the Popularity-Similarity-Optimization (PSO) model (size *N*, half of average degree *m*, temperature *T*), fixing the power-law degree distribution exponent $\gamma = 2.5$. For each PSO network we generated 1000 random networks using the null-models: Maslov-Sneppen (MS), Cannistraci-Muscoloni-sum (CM-sum) and Cannistraci-Muscoloni-prod (CM-prod). For the sake of clarity, the CM model reported in the main article is the CM-prod. The plots report, for each different parameter combination, the rich-club coefficient (non-normalized) of the PSO network and the mean rich-club coefficient for each null-model, averaged over the 1000 repetitions.